\documentclass[twocolumn,prb,nobibnotes,altaffilletter,amsmath,amssymb,amsfonts]{revtex4}
\usepackage[latin1]{inputenc}
\usepackage{graphicx}
\usepackage{amsmath}
\usepackage{latexsym}
\usepackage{epsfig}

\usepackage[latin1]{inputenc}
\usepackage{graphicx}
\usepackage{amsmath}
\usepackage{latexsym}
\usepackage{epsfig}

\begin{document}
\title{Predicting The Effective Temperature of a Glass}

\author{Nicoletta Gnan}
\email{ngnan@ruc.dk}
\affiliation{DNRF Center ``Glass and Time'', IMFUFA, Dept. of Sciences, Roskilde University, P.O. Box 260, DK-4000 Roskilde, Denmark}
\author{Claudio Maggi}
\affiliation{DNRF Center ``Glass and Time'', IMFUFA, Dept. of Sciences, Roskilde University, P.O. Box 260, DK-4000 Roskilde, Denmark}
\author{Thomas B. Schr{\o}der}
\affiliation{DNRF Center ``Glass and Time'', IMFUFA, Dept. of Sciences, Roskilde University, P.O. Box 260, DK-4000 Roskilde, Denmark}
\author{ Jeppe C. Dyre }
\affiliation{DNRF Center ``Glass and Time'', IMFUFA, Dept. of Sciences, Roskilde University, P.O. Box 260, DK-4000 Roskilde, Denmark}

\date{\today}
\begin{abstract}
We explain the findings by Di Leonardo {\it et al.} [Phys. Rev. Lett. \textbf{84}, 6054 (2000)] that the effective temperature of a Lennard-Jones glass depends only on the final value of the density in the volume and/or temperature jump that produces the glass phase. This is not only a property of the Lennard-Jones liquid,  but a feature of all strongly correlating liquids. For such liquids data from a single quench simulation provides enough information to predict the effective temperature of any glass produced by jumping from an equilibrium state. This prediction is validated by simulations of the Kob-Andersen binary Lennard-Jones liquid and shown not to apply for the non-strongly correlating monatomic Lennard-Jones Gaussian liquid.
\end{abstract}

\maketitle

Condensed matter is frequently found in out-of-equilibrium states. For example, for systems like supercooled liquids, dense colloids, spin systems, etc., the (off-equilibrium) glass state occurs naturally after cooling or compression from a state of thermal equilibrium. An \emph{effective temperature} describes the non-equilibrium properties of a glass, and the possibility of connecting the effective temperature with the observed violation of the \emph{fluctuation-dissipation theorem} \cite{FDT} (FDT) has opened new ways of inquiry \cite{Parisi,Sellitto,Kob, RDL,Jack}. In 2000 Di Leonardo {\it et al.} \cite{RDL} studied the off-equilibrium dynamics of the single-component Lennard-Jones (LJ) liquid (with a small many-body term added to the potential to prevent crystallization). This system was subjected to sudden temperature decreases at constant density (\emph{quenches}) as well as to sudden density increases at constant temperature (\emph{crunches}). From the violation of the FDT the effective temperature was determined. Surprisingly, it was observed that the effective temperature $T_\mathrm{eff}$ is independent of the particular path in the temperature-density plane crossing the glass transition line: $T_\mathrm{eff}$ depends only on the final density. In this Letter we demonstrate that the findings of Di Leonardo \emph{et al.} hold only for \textit{strongly correlating liquids} (defined below). We further argue and demonstrate that -  for this class of liquids - from a single quench simulation one can predict the effective temperatures for any off-equilibrium jump.

Reference \onlinecite{scl} documented the existence of a large class of liquids characterized by strong correlations between virial ($W\equiv PV-Nk_BT$) and potential energy ($U$) thermal equilibrium fluctuations at fixed volume, $\Delta W(t) \cong  \gamma\Delta U(t)$. Strongly correlating liquids have a hidden (approximate) scale invariance, which implies that they inherit many - but not all - of the scaling properties of liquids interacting via inverse power-law potentials. Strongly correlating liquids include van der Waals liquids like Lennard-Jones type liquids but not, e.g., hydrogen-bonding liquids. Strongly correlating liquids have curves in their phase diagrams -- ``isomorphs'' -- along which several static and dynamic properties are invariant \cite{IV}. These invariants derive from the fact that two microscopic configurations of two isomorphic state points, which trivially scale into one another, to a good approximation have proportional canonical probabilities. If the density is denoted by $\rho$ and the temperature by $T$, an isomorph is given by $\rho^\gamma/T={\rm Const.}$ The exponent $\gamma$ - which may be slightly state-point dependent - can be calculated from a quench simulation  utilizing the relation between the relaxing averages, $\langle W(t) \rangle \cong \gamma \langle U(t) \rangle + W_0$.

\begin{figure}
\begin{center}
\includegraphics[width=8.5cm]{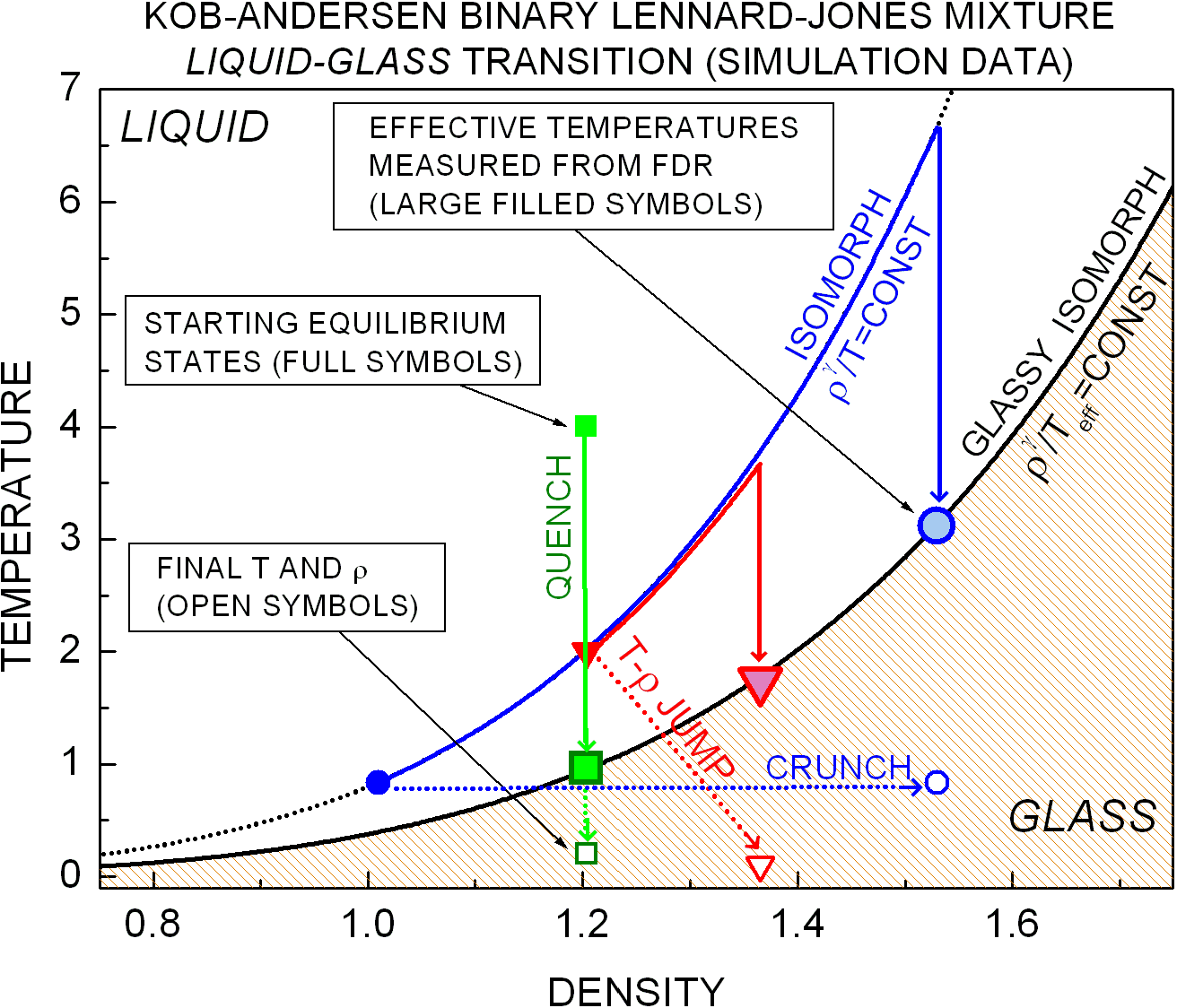}
\end{center}
\caption{Patterns followed by the KABLJ liquid in different off-equilibrium density and/or temperature jumps. Consider for example the case of a crunch (horizontal dotted line), where the system is densified at constant temperature. This transformation is equivalent to a quench (right-most vertical line) from an isomorphic state point having the final density of the crunch. Thus the $T_{\rm eff}$ (filled circle) is identical for these transformations. In all processes represented here the liquid undergoes a glass transition characterized by an effective temperature that can be measured from the fluctuation-dissipation relation (FDR) Eq. (\ref{eq1}). $T_{\rm eff}$ versus (final) density constitutes an isomorph, as discussed later in the text.}
\label{fig:F0}
\end{figure}

Because the canonical probabilities of scaled configurations belonging to the same isomorph are identical, a jump between two isomorphic state points takes the system instantaneously to equilibrium (property (\textit{i}))\cite{IV}. Moreover, jumps from isomorphic state points to the same final state point show identical aging behavior (property (\textit{ii})) \cite{IV}. In view of these properties the results of Di Leonardo {\it et al.} \cite{RDL} may be understood as follows. A crunch from density $\rho_1$ to density $\rho_2$ can be ideally decomposed into two parts (Fig. \ref{fig:F0}): First the system jumps \emph{instantaneously} from its initial state to the corresponding isomorphic state at the final density (i.e., the state which has the same $\rho^\gamma/T$ as the initial state), see  the blue curve of Fig. \ref{fig:F0}. This is an equilibrium state \cite{IV}. Thereafter the system begins to thermalize to the final temperature at constant density. If the crunch is made to a state with very high density,  the thermalization takes extremely long or maybe infinite time, and the effective temperature may be determined from the FDT violation as detailed below. In this way any crunch corresponds to a quench to the final density with  the same relaxation pattern. In particular, these two transformations should have identical FDT violation factors and identical effective temperatures.

These ideas should apply to any strongly correlating liquid, not just the single-component LJ system. To confirm this we simulated the standard Kob-Andersen binary Lennard-Jones (KABLJ) system \cite{BerthierKob,aux}. Following Di Leonardo {\it et al.} \cite{RDL} we subjected the KABLJ liquid to a number of instantaneous  quenches and crunches and calculated the effective temperatures from the  \emph{fluctuation-dissipation relation} (FDR). Recall that for off-equilibrium systems the FDR in $k_B=1$ units is \cite{Vulpiani,Kurchan,Crisanti,Cugliandolo,Cugliandolo2}

\begin{equation}\label{eq1}
 T\partial_{t'} \chi(t,t')= -X(t,t') \partial_{t'} C(t,t') .
\end{equation} 

\noindent Here $C=\langle A(t) B(t') \rangle$ is the equilibrium correlation function of the variables $A$ and $B$ in the unperturbed situation, the perturbing contribution to the Hamiltonian is $\delta H=-\epsilon B$, $\chi(t,t')=\langle A(t) \rangle/\epsilon|_{ \varepsilon \rightarrow 0}$ is the response of $A$ to an infinitesimal external field of magnitude $\varepsilon$ applied at time $t'<t$, and $X$ is the \emph{FDT violation factor}. This is unity at short times $(t-t')/t' \ll 1$ while $X<1$ in the long-time limit $(t-t')/t' \gg 1$. We chose as dynamic variables $A_{\mathbf{k}}(t)=N^{-1}\sum_j \eta_j \exp(i \mathbf{k}\cdot\mathbf{r}_j(t))$ and $B_{\mathbf{k}}(t)=N[A_{\mathbf{k}}(t)+A_{-\mathbf{k}}(t)]$, where the sum is extended to all $N$ particles of the system and $\eta_j = \pm 1$ is a random variable with zero mean. With this choice the correlation function studied is the self-intermediate scattering function, and $C(t,t')$ is dimensionless. 

For quenches to low enough temperatures, at long times an effective temperature of the slow degrees of freedom $T_{\rm eff}$ is associated with the FDT violation factor: $T_{\rm eff}=T/X$ \cite{Kurchan,Crisanti,Cugliandolo,Cugliandolo2,Leuzzi}.  The effective temperature reflects the slow structural rearrangements in the sense that the aging system behaves as if it were thermalized at $T_\mathrm{eff}$ \cite{Leuzzi}. We obtained $X$ by calculating the correlation function and the response function in the non-equilibrium regime by means of $X=X(t)=-\partial [T\chi(t,t')]/\partial C(t,t')|_t$, which applies at long times (note that the correct $X$ is found by taking this derivative at fixed $t$, not at fixed $t'$ \cite{Berthier}).

\begin{figure}
\begin{center}
\includegraphics[width=8.75cm]{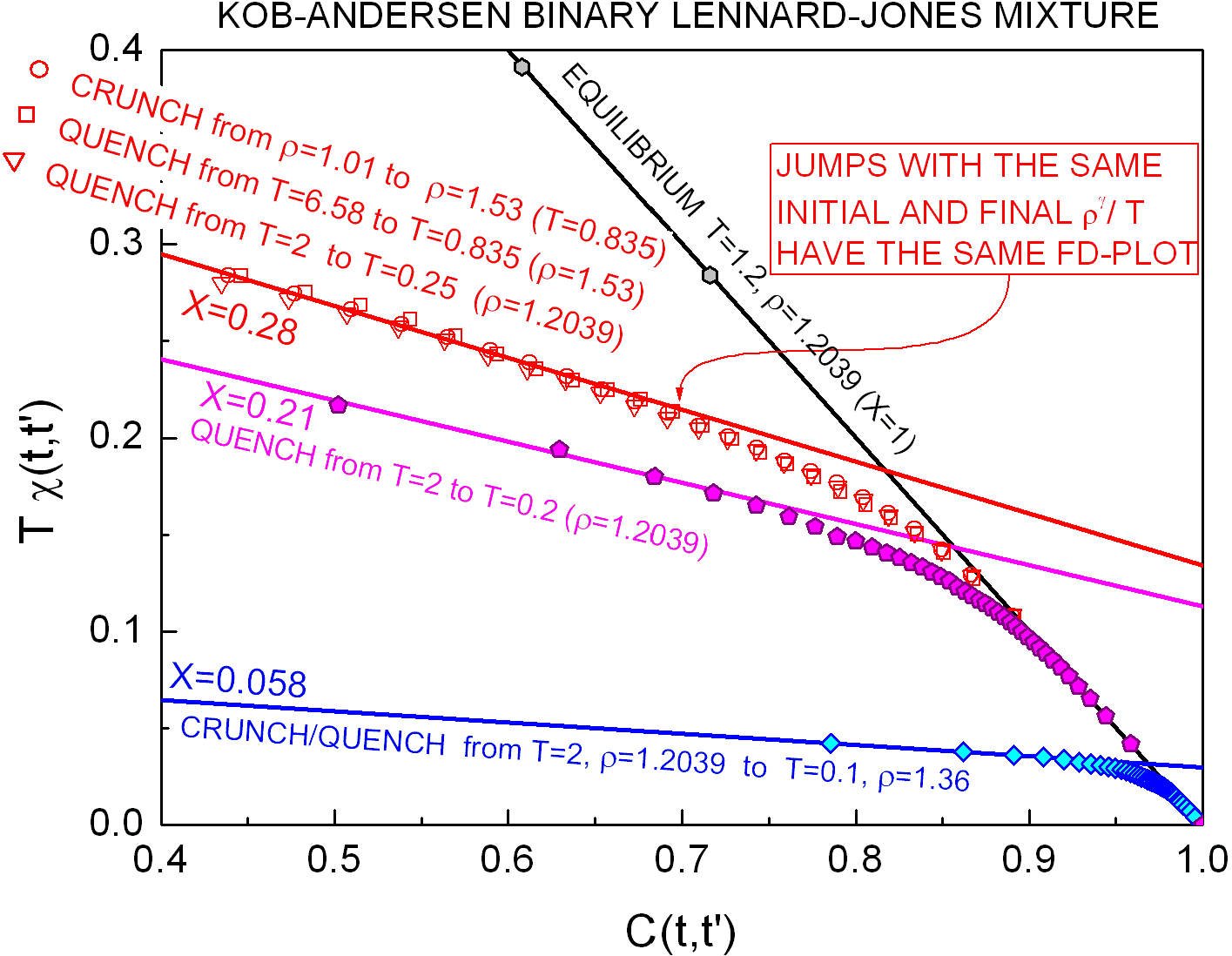}
\end{center}
\caption{Response vs. correlation function for several density/temperature jumps for the KABLJ system. All FD-plots have fixed $t=10^4$ (Monte Carlo steps) and $t'$ varying from $10^3$ to $10^4$. All functions plotted here have the same reduced k-vector \cite{aux} (referring to the final density) ({\large$\circ$}) and the same reduced microscopic time \cite{aux}. In the crunch ({\large$\circ$}) we set $|\mathbf{k}|=7.81$ corresponding to the reduced k-vector $|\tilde{\mathbf{k}}|=6.78$ (see \cite{aux} for details). The crunch ({\large$\circ$}) overlaps very well with the quench ($\Box$) that takes the system from an initial state isomorphic to the one of the crunch to the same final state. Note also the good superposition of the additional quench ({\footnotesize$\bigtriangledown$}) that takes the system from a state isomorphic to the initial state of ({\large$\circ$}) to a state isomorphic to the final one of ({\large$\circ$}). The full lines have the slopes predicted from the density scaling relation Eq. (\ref{eq:teffscal}) for $T_{\rm eff}$, adjusted only by a vertical shift to fit the data.} \label{fig:F2}
\end{figure}

\begin{figure}
\begin{center}
\includegraphics[width=8.75cm]{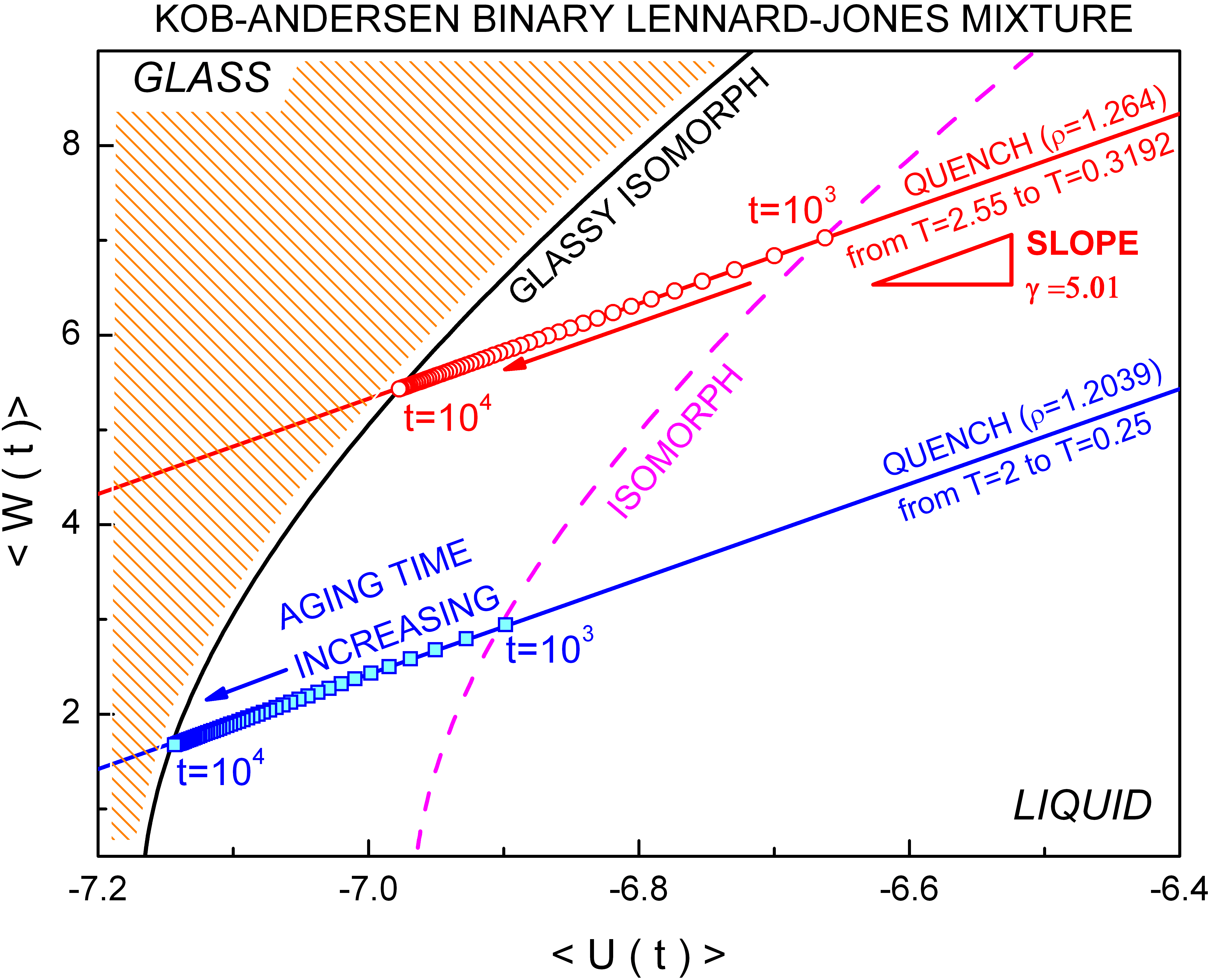}
\end{center}
\caption{Average virial versus potential energy per particle during aging in two quenches for the KABLJ system. These quenches are performed between states with the same initial and final $\rho^\gamma/T$. At each time the off-equilibrium states are connected by an isomorph (e.g., the dashed line). In these two jumps the FDT violation factor $X$ should be identical. The slope of this $\langle W(t) \rangle$ vs. $\langle U(t)\rangle$ plot is $\gamma$. The analytical equations used for drawing the isomorphs in the $W$-$U$ plot here are reported in Ref. \cite{IVv1}.}
\label{fig:F1}
\end{figure}

Recently Berthier introduced a new method for calculating the response without applying any external field to an off-equilibrium Monte-Carlo  simulation of the KABLJ \cite{Berthier}. We use his procedure and study the same dynamic variables for building the correlation function and the response as in Ref. \onlinecite{Berthier}. Figure \ref{fig:F2} shows the FD-plots for the KABLJ system during a number of temperature/density jumps. In Fig. \ref{fig:F2} we test the construction of equivalent crunches and quenches argued above: a crunch and a quench from initial isomorphic states (i.e., with the same $\rho^\gamma/T$) to the same final $T$ and $\rho$ (red circles and squares). Clearly, the crunch overlaps well with the quench since - as argued above - they follow the same aging pattern. The exponent $\gamma$  was estimated by a linear fit of the parameter plot  $\langle W(t) \rangle$ vs $\langle U(t) \rangle$ when the system is relaxing after a temperature jump from $T=2.55$ to $T=0.319$ at fixed $\rho=1.264$. The resulting value $\gamma=5.01$ is used throughout this paper. Fig. \ref{fig:F1} shows the linear relation that connects $\langle W(t) \rangle$ and $\langle U(t) \rangle$ when the system performs out-of-equilibrium jumps. 

Identical responses and correlations do not only appear when a strongly correlating liquid is taken from two isomorphic state to the same state point. Supplementing properties (\textit{i}) and (\textit{ii}), strongly correlating liquids have a third interesting aging property (\textit{iii}): For two jumps $(T_1,\rho_1)\rightarrow(T_2,\rho_2)$ and $(T_3,\rho_3)\rightarrow(T_4,\rho_4)$ between mutually isomorphic initial and final states (i.e., $\rho_1^\gamma/T_1=\rho_3^\gamma/T_3$ and $\rho_2^\gamma/T_2=\rho_4^\gamma/T_4$) the systems follow the same path in configuration space in reduced units (see \cite{aux}), and the dynamical equations governing the evolution of the particle trajectories are identical when cast in reduced units. Accordingly, the responses and correlations of two such jumps must be identical in reduced units. In Fig. \ref{fig:F2} we show the reduced-unit $C$ and $\chi$ of a quench between initial and final states that are isomorphic, respectively, to the initial and final states of the crunch described above (red triangles). The overlap between the functions is good. This concept is visualized in Fig. \ref{fig:F1}, which shows the variables $\langle W(t) \rangle$ vs $\langle U(t)\rangle$ in two jumps (with the same initial and final $\rho^\gamma/T$) connected by an isomorph at each time $t$ during the relaxation. 

A further consequence of property (\textit{iii}) is the following. Because the reduced-unit evolution of the dynamical properties is the same for the system in the two jumps, their FDR violation factors must also be identical, $X_2=X_4$. Combining this equation with $\rho_2^\gamma/T_2=\rho_4^\gamma/T_4$ and expressing $X$ via the effective temperature, we find $\rho_2^\gamma/T_{{\rm eff,2}}=\rho_4^\gamma/T_{{\rm
eff},4}$. This implies that

\begin{equation} \label{eq:teffscal}
\rho^\gamma/T_{\rm eff}=\mathrm{Const.}
\end{equation}
\noindent This equation identifies the dynamic glass transition curve in the ($T,\rho$) plane defined in terms of the FDR effective temperature with an isomorph. This is consistent with the findings of Ref. \onlinecite{RDL} and the standard way of defining the glass transition, because the standard glass line in the ($T,\rho$) plane is located where the equilibrium relaxation time reaches a certain (high) value of order the inverse cooling rate. For strongly correlating liquids an isomorph is also an ``isochronal'' curve along which the (reduced) relaxation time is constant \cite{IV}. Figure \ref{fig:F2} shows the slopes predicted by Eq. (\ref{eq:teffscal}) (lines); clearly the prediction is fulfilled. 

\begin{figure}
\begin{center}
\includegraphics[width=8.75cm]{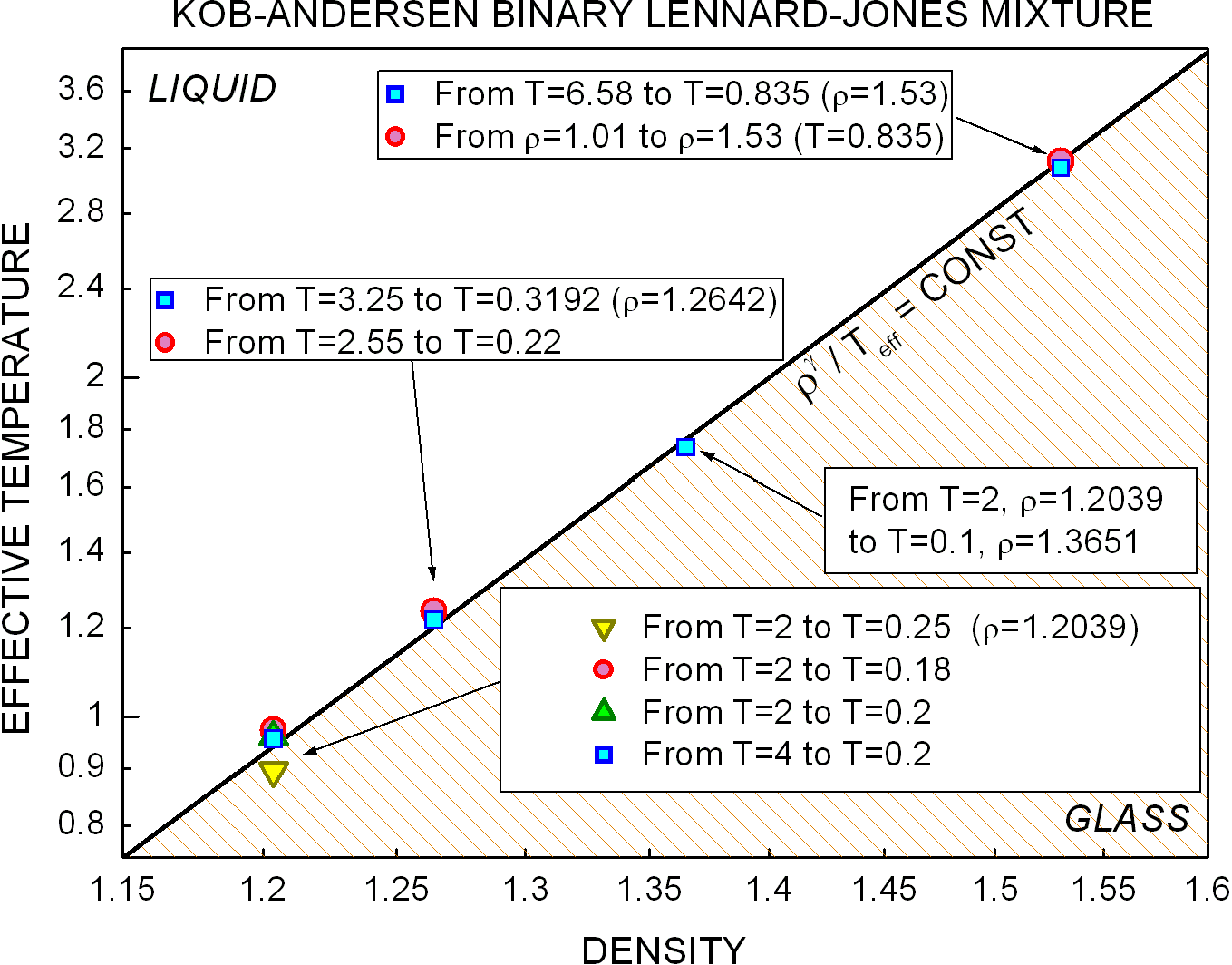}
\end{center}
\caption{Effective temperature as a function of density in several crunch and/or quenches (in double log-scale) for the KABLJ system. The effective temperature here is computed from the violation factor $X$, which is estimated as the slope that fits best the FD-plot (see Fig. \ref{fig:F2}) at points having $(t-t')/t' \geq 1$. The scaling exponent $\gamma$ is computed from potential energy-virial relaxation (see Fig. \ref{fig:F1}) as described in the text. The full line is the prediction of the density-scaling equation (\ref{eq:teffscal}) for $T_{\rm eff}$. }
\label{fig:F3}
\end{figure}

It is well known (see for example \cite{Kob} and \cite{Berthier}) that the effective temperature is independent of the initial and final temperature if the initial temperature is high (the system is in a warm liquid state) and if 
the quenching temperature is low enough (i.e., in the linear regime where $X=T/T_{\mathrm{eff}}$ with constant $T_\mathrm{eff}$). Equation (\ref{eq:teffscal}) predicts the effective temperatures for all state points of the phase diagrams, with the exponent $\gamma$ as well as the constant found from one single aging experiment. In Fig. \ref{fig:F3} we compare the $T_{\mathrm{eff}}$ measured in many crunches and quenches (not only involving isomorphic initial and final state points) with the prediction of Eq. (\ref{eq:teffscal}). The agreement is very good.

\begin{figure}
\begin{center}
\includegraphics[width=8.75cm]{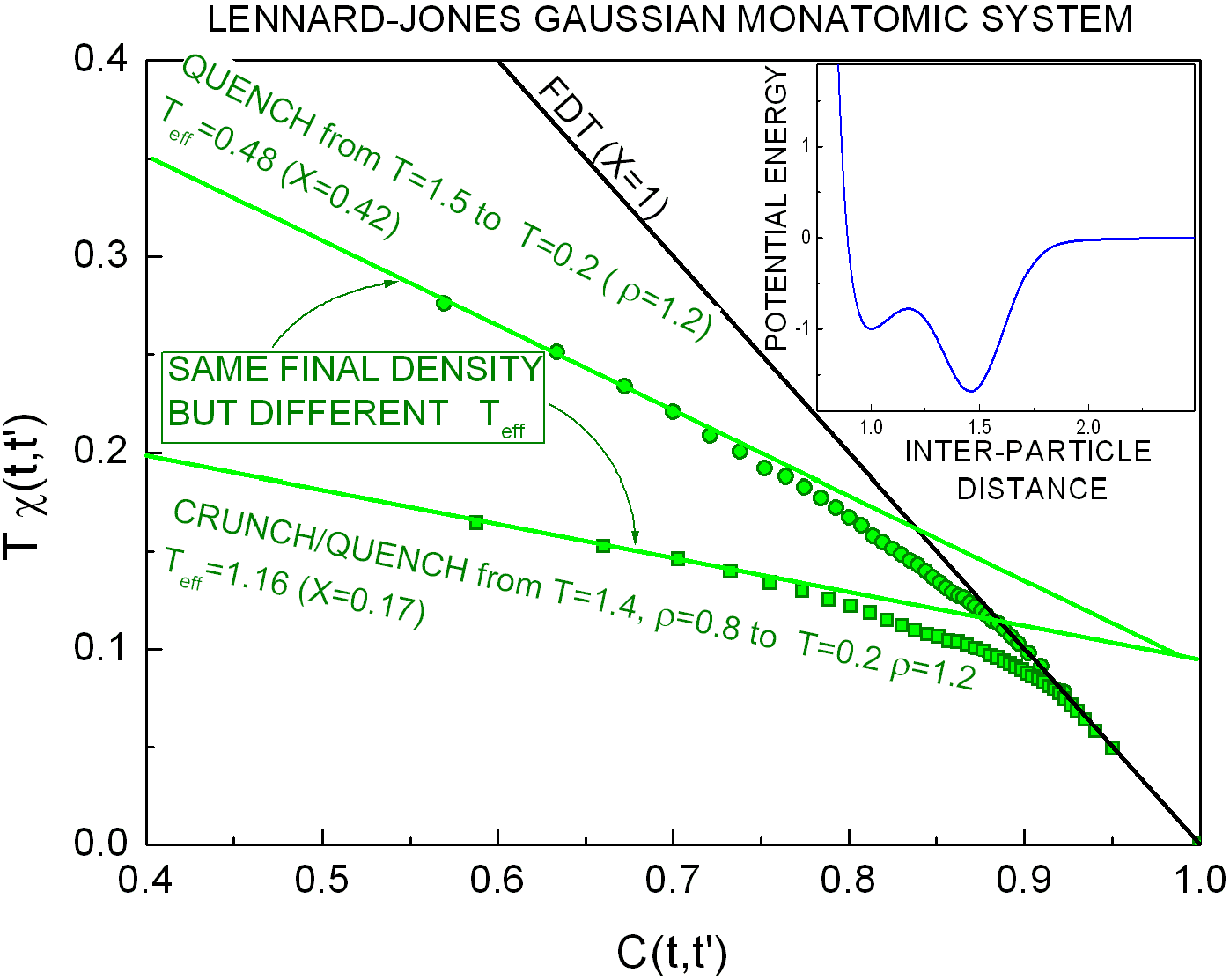}
\end{center}
\caption{FD-plot in two off-equilibrium transformations for the MLJG system at the same final density and temperature. The potential defining this model is shown in the inset. In this case the two effective temperatures are very different.}
\label{fig:F4}
\end{figure}

The above discussed simple aging properties are only expected to apply for liquids with isomorphs, i.e., strongly correlating liquids. To validate this we simulated the non-equilibrium dynamics of the \textit{monatomic Lennard-Jones Gaussian} (MLJG) model \cite{MLJG}. The pair-potential of the MLJG has an additional Gaussian attractive well compared to the LJ liquid (see the inset of Fig. \ref{fig:F4}); details about the model's potential and its glassy behavior can be found in Ref. \onlinecite{MLJG}. The MLJG liquid has $WU$ fluctuations which correlate less than 2\% at the state points studied here. As is clear from Fig. \ref{fig:F4} two jumps to the same final density lead to quite different effective temperatures. Thus this system provides a counterexample to the observation by Di Leonardo {\it et al.} that the effective temperature depends only on the final density.

We also investigated the relationship between the inherent state energies in aging and at the equilibrium for the KABLJ and the MLJG (see \cite{aux} for more details). We find that only for the strongly correlating liquid KABLJ can one interprete $T_\mathrm{eff}$ as an indicator of which part of the energy landscape is visted during aging (as suggested by Sciortino {\it et al.} in \cite{Sciortino}). 

In conclusion, the existence of isomorphs for strongly correlating liquids explains the previously reported result for the LJ liquid that the effective temperature depends only on the final density of any jump (when temperature and density are the externally controlled variables) \cite{RDL}. We presented simulations of the aging dynamics of another strongly correlating liquid (the KABLJ system), as well as simulations of aging of a liquid without strong virial / potential energy correlations (the MLJG liquid). For a strongly correlating liquid it is always possible to produce equivalent density/temperature transformations connected by the density-scaling relation. Moreover, for this class of liquids the effective temperature satisfies the density-scaling equation Eq. (\ref{eq:teffscal}). Since the exponent $\gamma$ and the constant of Eq. (\ref{eq:teffscal}) may be identified from a single quench simulation, the implication is that for a strongly correlating liquid the effective temperature of an arbitrary glass may be calculated from the results of a single jump simulation.

The center for viscous liquid dynamics \emph{Glass and Time} is sponsored by the Danish National Research Foundation (DNRF).

\section{Auxiliary Material}

\subsection{Simulation Details}

We simulate a binary mixture (80:20) of 1000 particles interacting via Kob-Andersen Lennard-Jones potential. Simulations are performed in the canonical (NVT) ensemble using the Metropolis Monte Carlo algorithm. Simulation parameters are taken from Ref. \onlinecite{BerthierKob}.

\subsection{Reduced Units in The MC Simulation}

The reduced length \cite{IV} is given by
$\tilde{l}=\rho^{1/3}l$. This defines the reduced k-vector
$\tilde{\mathbf{k}}=\rho^{-1/3}\mathbf{k}$ implying that, given a wave-vector
$\mathbf{k}_1$ at density $\rho_1$, the k-vector with equivalent
$\tilde{\mathbf{k}}$ at density $\rho_2$ is
$\mathbf{k}_2=(\rho_2/\rho_1)^{1/3}\mathbf{k}_1$. 

The microscopic time $\tau$ in the MC dynamics is set by the
maximum random step attempted by the particle $\delta_{\mathrm{max}}$, i.e.
$\tau=\tau(\delta_{\mathrm{max}})$. This is fixed to minimize the relaxation time of the
(equilibrium) correlation function at a given intermediate $T$ and $\rho$ (we
set $\delta_{\mathrm{max}}$ at the same value chosen in \onlinecite{Berthier,BerthierKob}
minimizing the decay-time of $C$ at $T=0.75$ and $\rho=1.2$). It follows that
two MC simulations (performed at $\rho_1$ and $\rho_2$ respectively) have the
same reduced time $\tilde{\tau}$ if $\tilde{\delta}_{\mathrm{max}}$ is the same, that is
when $\delta_{\mathrm{max}2}=(\rho_1/\rho_2)^{1/3}\delta_{\mathrm{max}1}$. However note that this
is minor adjustment of $\delta_{\mathrm{max}}$: in the density range explored (in our
simulations $\rho$ varies from 1 to 1.53) $\delta_{\mathrm{max}}$ changes at most 15 \%.

\subsection{Inherent States Energy: Equilibrium and Off-Equilibrium}

\begin{figure}
\begin{center}
\includegraphics[width=8.65cm]{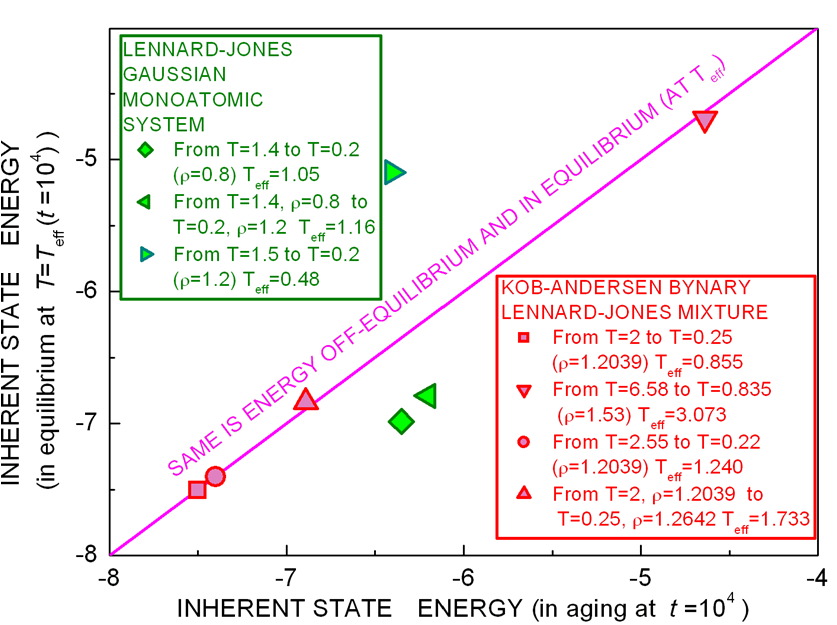}
\end{center}
\caption{Comparison between the (average) equilibrium inherent states energy (at
$T=T_{\mathrm{eff}}(t)$) and
the off-equilibrium inherent states energy (at time $t$) in several jumps in the glass (see legend).
For the KABLJ systems (red symbols) these two quantities coincide, indicating that the aging system at $t$
visits the same  inherent states of the system in equilibrium at the temperature
$T=T_{\mathrm{eff}}(t)$ given by the FDR. This is clearly not the case for the
MLJG systems (green symbols). The energy values of the MLJG are scaled by a factor 2.5
for an easier comparison with the KABLJ energies.}
\label{fig:F6}
\end{figure}

For a number of our simulations we evaluated the mean energy of the
inherent 
states (local minima in the 3N+1 dimensional potential energy landscape) visited at $t = 10^4$. 
In Fig. \ref{fig:F6} this is plotted against the mean inherent energy in
equilibrium 
when the system is simulated at $T=T_{\mathrm{eff}}$ of corresponding aging experiment. For the 
strongly correlating KABLJ system there is good agreement between these two quantities, 
consistent with the interpretation that $T_{\mathrm{eff}}$ indicates what part of the potential 
energy landscape is visited during the aging \cite{Sciortino}. In contrast, 
this interpretation is clearly not valid for the non-strongly correlating MLJG system, 
supporting the notion that strongly correlating liquids have simpler physics than liquids 
in general \cite{scl}.

\begin{thebibliography}{99}

\bibitem{FDT} J.-P. Hansen and I.R. Mc-Donald. Theory of Simple Liquids (Academic Press 2006)

\bibitem{Parisi} G. Parisi, Phys. Rev. Lett. \textbf{79}, 3660 (1997).

\bibitem{Sellitto} M. Sellitto, Eur. Phys. J. B {\bf 4}, 135 (1998).

\bibitem{Kob} W. Kob and J. L. Barrat, Eur. Phys. Lett. 46, 637 (1999).

\bibitem{RDL} R. Di Leonardo, L. Angelani, G. Parisi, and G. Ruocco. Phys, Rev. Lett. \textbf{84}, 6054 (2000).

\bibitem{Jack} R. L. Jack, M. F. Hagan and D. Chandler, Phys. Rev. E \textbf{76}, 021119 (2007)

\bibitem{scl} 
U. R. Pedersen, N. P. Bailey et al. , Phys. Rev. Lett. {\bf 100}, 015701 (2008); 
U. R. Pedersen, T. Christensen et al. Phys. Rev. E {\bf 77}, 011201 (2008);
N. P. Bailey, U. R. Pedersen et al., J. Chem. Phys. {\bf 129}, 184507 and 184508 (2008);
T. B. Schr{\o}der, U. R. Pedersen et al., Phys. Rev. E {\bf 80}, 041502 (2009).

\bibitem{IV} N.Gnan, T. B. Schr{\o}der, U. R. Pedersen, N. P. Bailey, and J. C. Dyre, J. Chem. Phys. { \bf 131} 234504  (2009).

\bibitem{BerthierKob} W. Kob and H. C. Andersen, Phys. Rev. Lett. {\bf 73}, 1376 (1994); L. Berthier and W. Kob, J. Phys.: Condens. Matter {\bf 19}, 205130 (2007).

\bibitem{aux} N. Gnan, C. Maggi et al. Auxiliary Material (2009).

\bibitem{Vulpiani} U. M. Marconi, A. Puglisi, L. Rondoni and A. Vulpiani, Phys. Rep. \textbf{461}, 111 (2008).

\bibitem{Kurchan} J. Kurchan, Nature \textbf{433}, 222 (2005).

\bibitem{Crisanti} A. Crisanti and R. Ritort, J. Phys. A \textbf{36}, R181 (2003).

\bibitem{Cugliandolo} L. F. Cugliandolo Slow Relaxations and nonequilibrium dynamics in condensed matter Course 7: Dynamics of Glassy Systems (Springer Berlin / Heidelberg 2004).

\bibitem{Cugliandolo2}  C. Chamon and L. F. Cugliandolo, J. Stat. Mech. P07022 (2007).

\bibitem{Leuzzi} L. Leuzzi J. Non-Cryst. Solids \textbf{355}, 686 (2009).

\bibitem{Berthier} L. Berthier, Phys. Rev. Lett. {\bf 98}, 220601 (2007).

\bibitem{IVv1} N.Gnan, T. B. Schr{\o}der et al. arXiv:0905.3497v1.

\bibitem{Sciortino} F. Sciortino and P. Tartaglia Phys. Rev. Lett. \textbf{86}, 107 (2001)

\bibitem{MLJG} V. Van Hoang and T. Odagaki, Physica B \textbf{403}, 3910 (2008).

\end{thebibliography}
\end{document}